\documentclass[conference]{IEEEtran}
\IEEEoverridecommandlockouts
\usepackage{cite}
\usepackage{amsmath,amssymb,amsfonts}
\usepackage{algorithmic}
\usepackage{graphicx}
\usepackage{textcomp}
\usepackage{xcolor}
\usepackage{lipsum,comment}
\usepackage{array}
\def\BibTeX{{\rm B\kern-.05em{\sc i\kern-.025em b}\kern-.08em
    T\kern-.1667em\lower.7ex\hbox{E}\kern-.125emX}}

\def\ie{\textit{i.e.}}

\begin{document}
\title{ROSE: A RObust and SEcure DNN Watermarking}

\author{
\IEEEauthorblockN{Kassem Kallas and Teddy Furon}
\IEEEauthorblockA{Centre Inria de l'Université de Rennes, France \\
firstname.lastname@inria.fr}
%\IEEEauthorblockN{2\textsuperscript{nd} Given Name Surname}
%\IEEEauthorblockA{\textit{INRIA} \\
%Rennes, France \\
%email address or ORCID}
}

\maketitle

\begin{abstract}\label{abstract}
Protecting the Intellectual Property rights of DNN models is of primary importance prior to their deployment. So far, the proposed methods either necessitate changes to internal model parameters or the machine learning pipeline, or they fail to meet both the security and robustness requirements. This paper proposes a lightweight, robust, and secure black-box DNN watermarking protocol that takes advantage of cryptographic one-way functions as well as the injection of in-task key image-label pairs during the training process. These pairs are later used to prove DNN model ownership during testing. The main feature is that the value of the proof and its security are measurable. The extensive experiments watermarking image classification models for various datasets as well as exposing them to a variety of attacks, show that it provides protection while maintaining an adequate level of security and robustness.
\end{abstract}


\section{Introduction}\label{sec:introduction}
Due to their exceptional performance, Deep Neural Networks (DNNs) are being adopted and deployed in a wide range of real-world applications. They become the de facto standards for a wide range of computer vision tasks, including image classification. 
DNN model training is a time-consuming process that necessitates large datasets, significant computational resources, and expertise in optimizing DNN parameters and topology. Furthermore, commercial deployment and model sharing are critical to the advancement of DNNs. Big companies like Google, Apple, and Facebook, for example, have already implemented DNNs in products that we use on a daily basis \cite{SurveyDeployingDNN2022}. Therefore, protecting the ownership of DNNs is critical, as they are regarded as valuable industrial assets.
%an Intellectual Property (IP) of the Owner. 

The protection of Intellectual Property involves two parties: the claimed Owner provides evidence to a Verifier entity that a given deployed model is its ownership.
The literature shows that watermarking DNNs in a robust manner is technically possible. Removing the watermark from a stolen model is as demanding as training a new network from scratch in terms of computational resources and datasets. This justifies the use of DNN watermarking as the main tool for ownership protection.

The first focus of this paper is not the robustness of DNN watermarking but \emph{the value of the proof of ownership}.
A high detection score has no value if watermarking is not framed by a protocol. 
We propose to measure the value of the proof by computing the $p$-value, which is the probability that a randomly picked secret key produces a score higher than the one produced by the secret key of the claimed Owner.
This allows to quantify the size of the secret key to raise the value of the proof enough so that it removes any reasonable doubt in the mind of the Verifier.

Secondly, the Verifier should assume by default that the claimed Owner is an Usurper. A true Owner first generates a secret key and then watermarks his model, whereas the Usurper first steals a model and then finds a posteriori a pseudo secret key giving a watermark detection. We state that the Verifier should quantify the security level by gauging the amount of \emph{work} for finding such a pseudo secret key.

In the end, the Verifier grants ownership if and only if both \emph{work} and \emph{value of the proof} are large. The second contribution is a binding mechanism that makes the value of the proof linear with the size of the key and the work exponential with the value of the proof. This means that finding a posteriori a pseudo secret key is NP hard with the size of the key.
This scheme does not weaken the robustness of the watermark as experimentally tested on several datasets.

\section{Background and Prior Art}\label{sec:background}
This section reviews the main classes of the algorithms to watermark DNN in the field of image classification. 

%Many of the DNN watermarking classifications criterias are naturally inherited from multimedia watermarking domain while, some others are particular to DNN domain and cannot be applied vice versa. A taxonomy that discusses the dissimilarities between multimedia watermarking and DNN watermarking can be found in \cite{FourchallengesandFuneral2021}.

\subsection{Classification}
DNN watermarking can be classified into white-box or black-box setting.
In the white-box case, the watermark is embedded directly into the DNN weights and may be recovered by performing a particular secret-keyed function of the weights or the activation maps.
%This is also known as static watermarking.
On the other hand, only the network output is accessible in the black-box setting. 
It looks after the network behavior and examines the network output to particular inputs designed for the watermark detection task.
These inputs injected in the training set are responsible to change the DNN behavior.
They are the Owner's secret key. These inputs are called \textit{entangled} if they belong to the same primary task that the DNN is supposed to performed.
%, otherwise, we call them \textit{non-entangled}.

Another class is the distinction between robust and fragile watermarking.
A robust watermark survives post-processing that may be performed intentionally to hide the host DNN. 
%for instance, fine-tuning, pruning, weights quantification.

This paper concerns black-box robust DNN watermarking with entangled inputs.
The watermark is embedded into the host DNN at training time or transfer learning.
In the first case, the watermark is embedded while training the model from scratch. In the second case, the feature part is loaded from a pre-trained model and the decision/fully-connected part is changed and re-initialized to fit to the transfer learning task and to embed the watermark.
%. Then, a retraining is performed using the new dataset we transfer the learning to.
%The method proposed in this paper and explained in section \ref{sec:proposedmethod} falls into black-box, zero-bit, dynamic watermark with entangled images category.

\subsection{DNN Watermarking Attacks}
Transfer learning, model pruning and weight quantization are the most common post-processing for measuring the robustness of DNN watermarking.

Pruning and weight quantization are methods reducing the network memory footprint. Pruning sets to zero a fraction of the model weights, while quantization casts the DNN weights onto smaller floating point or integer representations.
%These techniques heavily modify the network internal parameters.
They can play the role of an attack removing the watermark only if they do not decrease the accuracy of model too much since
the primary goal of the attacker is to get a working classifier.

Transfer learning applies the knowledge learned by a DNN on one task to a new similar task.
An example is to transfer the knowledge of a well-trained model on ImageNet~\cite{Imagenet2009} to the CIFAR10 classification problem~\cite{cifar102009}. Transfer learning typically involves replacing the decision part of the CNN (\ie\ fully connected layers) with new layers adequate for the new task.
A particular case is fine-tuning in which a trained DNN is re-trained with a lower learning rate and without decision layers replacement or re-initialization.
%In the following we will refer to the explained techniques as attacks and post-processing, alternatively.

\subsection{Prior Works}\label{sec:priorwork}
The first attempt of DNN watermaking dates back to 2017, by Uchida et al. in~\cite{uchida2017}. This is a white box setting where the watermark signal is directly embedded into the network weights of the host DNN.
Paper~\cite{SpreadDitherWatermark2021} increases the watermark payload by using informed coding, in particular Spread-Transform Dither Modulation.
%Chen et al. adds a traitor tracing mechanism that has anti-collusion capabilities in~\cite{Deepmarks2019}. A first watermark is embedded for IP protection and then, a second watermark carrying the user’s code is added prior to its release.
%In their algorithm, first they select a convolutional layer into which the watermark will be embedded. Then, the weights tensor of the layer is flattened after taking the average of the weights in the filter-dimension. The DNN is then trained by using an additional loss term ensuring that guarantee the watermak bits embedding and extraction from the weights tensor.

%In \cite{Deepsigns2019}, the authors proposed two solutions for DNN watermarking, one in a white-box setting and the other in black-box (multi-bit and zero-bit). In the multi-bit case, the N bits watermark is enforced into the activation maps by making the maps following a Gaussian Mixture Model (GMM). The embedding is performed using a modified loss function and input key images. In the zero-bit case, they select randomly pairs of images and labels from the training set as key. The key pairs are selected among the misclassified examples by the original non-watermarked model. First a non-watermarked model is trained, then the model is fine-tuned on misclassified key image-label pairs.

As for the black-box setting, the first work exploits backdooring to watermark a DNN~\cite{backdoorWM2018}.
The idea is to train the DNN such that it keeps a memory of some unusual pairs of input and label.
Unusual means that no other model would predict these labels for these inputs.
These inputs are called backdoors, watermarked inputs, secret keys, or triggers in the literature. They are synthetic~\cite{backdoorWM2018}, adversarial~\cite{Adversarialfrontier2020}, or benign inputs with visible~\cite{ProtectingIntellectualProperty2018} or invisible~\cite{Backdoorsignatureembeddedsystems2018} overlay.
The authors of~\cite{ProtocolVerification} warn that these kinds of triggers are distinguishable which allows the suspect to escape verification. This threat is indeed demonstrated in~\cite{ExponentialWeighting} which recommends to use a fraction of randomly picked benign training samples.
The labels of these triggers are usually randomly picked among the class set.
Any model learned for the same classification problem is unlikely to predict the same labels.

All the above-mentioned works demonstrate that watermarking does not spoil the accuracy of the model and that the watermark is robust against attacks (except~\cite{Backdoorsignatureembeddedsystems2018}). Yet, security has different definitions in the literature.
The authors of~\cite{FourchallengesandFuneral2021} make the analogy with classic media watermarking where security is defined as the inability for the attacker to estimate the secret key.
Some papers already prevent the disclosure of the secret trigger inputs by inversion of the watermarked model~\cite{ProtectingIntellectualProperty2018} or even by a semi-honest Verifier~\cite{ProtocolVerification}.
On the contrary, paper~\cite{Adversarialfrontier2020} relies on the steganographic definition of security where the attacker is should not be able to detect the presence of the watermark in a model.

\subsection{The Forgotten Funeral}
The analogy with classic watermarking~\cite{FourchallengesandFuneral2021} forgets one flaw (or funeral, to use the same metaphor as its authors) which is the threat of an Usurper claiming the ownership of a model. A given secret key raising a positive watermark detection on a given media or model has little value for proving ownership.
First, watermark detection has only statistical guarantees. If the probability of false alarm is $p$, then an attacker needs to sample in the order of $1/p$ keys to find one triggering a positive detection on a given media or model.
Essentially, a positive detection does not show that a key has been generated first and then the media or model has been watermarked. As known for a long time in classic watermarking, an Usurper may find a way to generate a posteriori a key suitable for a given model (more efficiently than by random key sampling), so that ``\textit{anyone can claim ownership of any watermarked image}'' ~\cite{Resolvingownerships}. For black-box DNN watermarking, targeted adversarial examples is a means to forge inputs resulting in any given prediction. This crafts a posteriori unusual pairs of input / label that may play the role of triggers. 

As far as we know, very few works in DNN watermarking consider the threat of an Usurper.
Paper~\cite{Backdoorsignatureembeddedsystems2018} proposes a secure protocol for generating the triggers and their labels, but robustness is not demonstrated at all. This is an issue since the triggers are generated by adding a very fragile overlay modifying very few pixels on some training images. According to our test, this watermark is not robust to the appending of a JPEG compression in front of the classifier.  
The authors of~\cite{backdoorWM2018} propose to bind triggers and their labels with a cryptographic commitment scheme, but strangely enough, adversarial examples are not considered: their claim ``\textit{this method makes back-propagation based attacks extremely hard}'' is not backed up by any experiment.

\def\real{\mathbb{R}}
\def\C{\mathcal{C}}
\def\S{\mathcal{S}}
\def\SS{M}%{S^\prime}
\def\ss{m}%{s^\prime}
\def\DT{\mathcal{D}_{train}}
\def\Pr{\mathbb{P}}
\def\defi{:=}
\def\wF{\omega_F}
\def\wH{\omega_H}
\def\TA{\textbf{acc}}
\def\rec{\textbf{rec}}
\section{Proposed Method}\label{sec:proposedmethod}
We propose a blind black-box watermarking which enforces the lessons of~\cite{Resolvingownerships}: i) the Owner cannot freely choose a secret key, ii) the Verifier must be convinced that the secret key has not been forged a posteriori.
We propose two metrics: a quantity gauging how convincing the proof of the Owner is, and the security level measuring the amount of work necessary for the Usurper to convince the Verifier.

We assume that crafting targeted adversarial examples with invisible and non detectable perturbation is feasible.
The cost of one network inference is denoted $\wF$, while the cost of one gradient computation is set to $2\wF$ (thanks to back-propagation). White-box adversarial attacks usually make a gradient descent over $t\approx 100$ iterations, so that the cost of forging one adversarial image is $2t\wF$. The cost of one hash computation is denoted $\wH$. 
This section explains the proposed method by gradually raising the security level.

\subsection{Level 0}
Consider the training set $\DT$ composed of $n$ pairs of inputs and labels $\{(x_i,y_i)\}_{i=1}^n\subset\real^d\times\C$ with $\C$ the set of classes $\{0,1,\ldots,c-1\}$.
% The internal parameters of the network are not necessarily accessible to the DNN trainer neither the loss function used in the optimization process.
Prior to learning, a small set of $s$ training inputs are secretly selected as triggers: up to a permutation, say these are the first training samples: $\{x_i\}_{i=1}^s$. Their labels are replaced by $s$ classes $\{\tilde{y}_i\}_{i=1}^s$ randomly picked in $\C$. This is likely to create a set of unusual pairs $\S=\{(x_i,\tilde{y}_i)\}_{i=1}^s$.
We force the Owner to use a secret key as the seed of the pseudo-random generator sampling these classes.
To sustain the accuracy of the model, the triggers represent a fraction of the training set: $s/n\ll 1$.
Then, training is done as usual and we assume that the learned model $F$ keeps a memory of almost all these pairs.

The Owner sends the triggers and the secret key to the Verifier which re-generates their labels.
The watermark detection amounts to verifying that $\tilde{y}_i=F(x_i)$ for most pairs in $\S$, say at least $\ss$ matches.
%As it is, there is no way to quantify how unusual are these pairs. For instance, if the classification problem is hard (like ImageNet), the accuracy of any model is far from perfect. Therefore, there exist plenty of misclassified inputs which could play the role of a would-be trigger. To solve this issue, the Owner has a secret key that is used as the seed of the pseudo-random generator sampling the labels $\{\tilde{y}_i\}_{i=1}^S$.
We propose to measure how unusual are these matching pairs by the rarity $R$ defined as a logarithmic function of the p-value: Selecting another key, the number of matches is a random variable $M$ and the p-value is the probability that at least $\ss$ pairs match among $\S$:
\begin{equation}
\label{eq:Rarity}
    R \defi - \log_2(\Pr(\SS\geq \ss))\quad \mbox{in bits}.
\end{equation}
For instance, if $R = 50$ bits, it means that the probability to accidentally find a secret key yielding such a large number of matches is $2^{-50}$. This is a very rare event giving evidence that the claim of ownership is correct. 

Picking a key at random, the event $\tilde{y}_i=F(x_i)$ occurs with probability $1/c$, so that the number $M$ of matches follows a binomial distribution $\SS\sim\mathcal{B}(s,1/c)$.
The true expression of this probability is
\begin{equation}
\label{eq:BetaInc}
    P(\SS\geq \ss) = I_{1/c}(\ss,s+1-\ss),
\end{equation}
where $I_x(a,b)$ is the regularized incomplete beta function.
A more workable expression is the Hoeffding inequality:
\begin{equation}
    P(\SS\geq \ss) \lesssim e^{-2s(r-1/c)^2},
\end{equation}
where $r \defi \ss/s$, which can be though as the watermark recovery rate (see Sect.~\ref{sec:experimentsResults}).
In other words, if the Verifier accepts the proof of ownership when the rarity is larger than $R$ bits, then the number of images to submit is:
\begin{equation}
\label{eq:sLinearR}
    s \gtrsim R \frac{\log(2)}{2(r-1/c)^2}.
\end{equation}
This is a decreasing function of the watermark recovery rate.
%As an example, for MNIST dataset in section \ref{sec:experimentsResults} with SHA1 (res. $N=40$), we can see that the probability for the DNN Owner to guess the correct labels of the key samples is 0.825, while the attacker cannot do better than random guesses, so the probability that an attacker will correctly guess the sequence $Y_h$ is $1e^{-40}$. Please keep in mind that the number of key images $N$ cannot be increased arbitrarily, such as by using a hash function with an arbitrarily long output stream like Skein \cite{skeinHash2010}, Keccak \cite{KeccakHashFucntion2014}, or others. The reason for this is that if the condition $\alpha \lll 1$ is not met, it may corrupt the original task for which the DNN was trained, causing the test accuracy on non-key examples to deteriorate, rendering the host DNN useless.

As for the security, an Usurper selects $s$ inputs and picks a key to generate $s$ random labels. Then he makes them match with a targeted adversarial attack. The work for this attack equals $W = 2st\wF$.
Since $s$ is linear with rarity $R$~\eqref{eq:sLinearR}, the work is then rather low as it is only linear with the rarity.
%\begin{equation}
%    W = 2st\wF \approx t R\log(2)\left(1-1/c\right)^{-2}\wF.
%\end{equation}

\subsection{Level 1}
The next level adds another constraint on the generation of the label of the triggers. The labels are imposed by a one-way function of the trigger. Let $H$ be a cryptographic hash function mapping a string into the set of integers $\{0,1,\ldots, 2^b-1\}$, where $b$ is the size of the binary hash.
This hash function is parametrized by the Owner's secret key $sk$.
We impose that $\tilde{y}_i = H(x_i;sk)\% c$, where $\%$ is the modulo operator. For $b$ large enough, this makes the labels uniformly distributed over $\C$ as in the previous scheme.
The claimed Owner sends the triggers and the secret-keyed hash function to the Verifier who grants ownership of model $F$ if $F(x_i)=H(x_i;sk)\% c$ for at least $\ss$ inputs of $\S$. The analysis of the rarity is the same as for Level 0.

At first sight, this increases the security level. Say $sk'$ is the secret key of the Usurper. The probability of forging one adversarial example $x$ which complies with the rule $F(x)=H(x;sk')\% c$ values $1/c$. The number of forgeries to get at least $\ss$ matches is distributed as a geometric distribution. On expectation, the work is now $W = 2tsc\wF$. 

This is indeed less that than: the Usurper first crafts one adversarial example with target class $y$ and then flips the Least Significant Bit of one random pixel until $H(x;sk')\%c=y$.  It is highly likely that one LSB flip does not modify the prediction of $F$ while it completely changes the hash value. This takes on expectation $c$ hash computations so that it decreases the amount of work to $W = s(2t\wF + c\wH)$. 

\subsection{Level 2}
Our final proposal is to hash jointly the $s$ triggers to generate their labels.
One simple implementation can be to first compute $h = H(x_1||\ldots||x_s;sk)$ and to use $h$ as the seed of a pseudo-random generator producing integers in $\{0,\ldots,2^b-1\}$, which are mapped to classes thanks to a modulo operation.
Again, randomly picking a secret key $SK$, the hash function $H(\cdot;SK)$ produces uniformly distributed labels in $\C$ so that the analysis of rarity $R$ remains the same.

As for the security, the Usurper first prepare $s$ adversarial examples with random target classes, modifies some LSB, computes the hash and the labels, and repeats until at least $\ss$ matches are observed.
Changing one LSB of one trigger modifies the label of all the triggers.
On expectation, the work now equals:
\begin{equation}
    W = 2st\wF + \frac{s\wH}{\Pr(\SS\geq \ss)} = s (2t\wF + 2^R\wH).
\end{equation}
The work is now much bigger as it is an exponential function of the rarity.

\section{Experimental results}
\label{sec:experimentsSetup}
\label{sec:experimentsResults}
\subsection{Setup}
The evaluation of the method uses MNIST, Fashion MNIST, CIFAR10 datasets plus ImageNet for transfer learning, and off-the-shelf CNN network architectures (see App.~\ref{sec:App}).
We divide the training dataset into three parts: 80\% for training, 10\% for validation, and 10\% for fine-tuning in all of the experiments.
The numbers of triggers is $s\in\{40, 64, 128\}$.

%The hash functions SHA1, SHA256, and SHA512 have output streams of length $\delta \in \{160, 256, 512\}$ corresponding to the number of key images $N \in \{40, 64, 128\}$.

As for the attacks, we consider pruning, weight quantization, JPEG compression and fine-tuning. We prune the entire DNN weights (or just the fully connected layers) with rate $k\in(0,1)$ by setting randomly selected weights value to zero.
Weight quantization can be done in three different ways. The weights are quantized into integers in dynamic range quantization (Dyn. Quant.) or in full integer quantization (Full Int. Quant.), or converted to Float16 format (Float16 Quant.). These operations reduce the size of the DNN and speed up querying time. Dynamic range quantization, for example, reduces the size by 4 while increasing the inference speed by 2 to 3. JPEG compresses and decompresses the input image before forwarding it to the CNN. The quality factor ranges from 55 to 100, with a step of 5.
Fine-tuning uses the same algorithm than training (see Sect.~\ref{sec:App}) but with a learning rate of $1e-5$, for 30 epochs and a batch size of 64.
In total, the robustness of the proposed method is tested against 30 attacks. We measure the test accuracy \TA\ and the watermark recovery rate \rec.

\subsection{Baseline Performance}
\label{subsec:BaselinePerformance}
The first conclusion is that increasing the number of triggers $s$ causes a minor drop in test accuracy on the classification task when comparing to non-watermarked model performance.
Starting with MNIST, the loss is 0.05 percent point with $s = 40$, 0.18 pp with $s = 64$, and 0.24 pp with $s = 128$.
The drop in test accuracy as $s$ increases is also negligible for Fashion MNIST.
CIFAR10 and transfer learning scenarios have a slightly higher drop:
the biggest loss is 0.66~pp with the former with $s=64$, 0.9~pp for the latter with $s=128$.
%Finally, based on the above findings on various datasets, the fidelity requirement in section \ref{sec:background} is met in all cases.

Secondly, the proposed method achieves decent results in terms of recovering the labels of the triggers.
The watermark recovery for MNIST ranges from 82.5\% for $s = 40$ to 87.5\% for $N = 128$.
This corresponds to a rarity $R$ of 97 to 308 bits, according to~\eqref{eq:Rarity}.
Fashion MNIST and CIFAR10 achieve a higher watermark recovery rate of 90.6\% for the former with $s=128$ and 92.5\% for the latter when $s=40$. Among all the cases, transfer learning has the highest watermark recovery performance accuracy, as evidenced by the fact that the \rec\ metric never falls below 90\% for any value of $s$. This again makes a rarity $R$ ranging from 103 to 332 bits.
These are large amounts of rarity which are extremely convincing for the Verifier even with only 40 triggers.

\newcommand{\mTc}[1]{\multicolumn{2}{c|}{#1}}
\newcommand{\mOc}[1]{#1}

\begin{table*}[]
\caption{Accuracy \TA, watermark recovery rate \rec\ against attacks, and the corresponding range of Rarity $R$ in bits}
\label{tab:PerformanceSecureHash}
\centering
%\resizebox{\textwidth}{!}{%
%\begin{tabular}{|cl||cl|cl|cl|cl|cl|}
\begin{tabular}{|cl||cc|cc|cc|cc||c|}
\hline
\multicolumn{2}{|c||}{} & \mTc{\textbf{Fine-Tune}}  & \mTc{\textbf{Dyn. Quant.}} & \mTc{\textbf{Full Int. Quant.}} & \multicolumn{2}{c||}{\textbf{Float16 Quant.}} & \mOc{\bf{Rarity}}\\
%\cline{3-12} 
\multicolumn{2}{|c||}{\textbf{Dataset $\backslash$ Nb. triggers}} & \mOc{\TA} & \mOc{\rec} & \mOc{\TA} & \mOc{\rec} & \mOc{\TA} & \mOc{\rec} & \mOc{\TA} & \mOc{\rec} & \bf{$R$ in bits}\\
\hline

\multicolumn{1}{|c|}{} & $s=40$ & \mOc{99.3} & {\bf 82.5} & \mOc{99.3} & {\bf82.5} & \mOc{99.3} & {\bf82.5} & \mOc{99.3} & {\bf82.5} & {\bf 86--86}\\
%\cline{2-12} 
\multicolumn{1}{|c|}{\textbf{MNIST}} & $s=64$ & \mOc{99.1} & {\bf89.1} & \mOc{99.1} & {\bf89.1} & \mOc{99.2} & {\bf89.1} & \mOc{99.2} & {\bf89.1} & {\bf167--167}\\
%\cline{2-12} 
\multicolumn{1}{|c|}{} & $s=128$ & \mOc{99.1} & 88.3 & \mOc{99.1} & {\bf87.5} & \mOc{99.1} & {\bf87.5} & \mOc{99.1} & {\bf87.5} & {\bf308}--320\\
\hline

\multicolumn{1}{|c|}{} & $s=40$ & \mOc{91.8} & {\bf85.0} & \mOc{91.7} & {\bf85.0} & \mOc{91.5}   & {\bf85.0} & \mOc{91.8} & {\bf85.0} & {\bf91--91}\\
%\cline{2-12} 
\multicolumn{1}{|c|}{\textbf{Fashion MNIST}} & $s=64$ & \mOc{91.9} & 89.1 & \mOc{91.9} & {\bf87.5} & \mOc{92.0} & {\bf87.5} & \mOc{91.7} & {\bf87.5} & {\bf155}--167\\
%\cline{2-12} 
\multicolumn{1}{|c|}{} & $s=128$ & \mOc{91.7} & {\bf89.8} & \mOc{91.9} & 90.6 & \mOc{92.0} & 90.6 & \mOc{91.8} & 90.6 & {\bf326}--332\\
\hline

\multicolumn{1}{|c|}{} & $s=40$ & \mOc{83.2} & {\bf92.5} & \mOc{83.4} & {\bf92.5} & \mOc{83.4} & {\bf92.5} & \mOc{83.4} & {\bf92.5} & {\bf110--110} \\
%\cline{2-12} 
\multicolumn{1}{|c|}{\textbf{CIFAR10}}& $s=64$ & \mOc{83.4} & {\bf85.9} & \mOc{83.2} & 87.5 & \mOc{83.2} & 87.5 & \mOc{83.1} & 87.5 & {\bf149}--155\\
%\cline{2-12} 
\multicolumn{1}{|c|}{} & $s=128$ & \mOc{84.0} & 90.6 & \mOc{83.3} & {\bf89.8} & \mOc{83.4} & {\bf89.8} & \mOc{83.3} & {\bf89.8} & {\bf326}--332\\
\hline

\multicolumn{1}{|c|}{} & $s=40$ & \mOc{85.1} & {\bf92.5} & \mOc{85.9} & {\bf92.5} & \mOc{86.0} & {\bf92.5} & \mOc{86.0} & {\bf92.5} & {\bf110--110}\\
%\cline{2-12} 
\multicolumn{1}{|c|}{\textbf{Transfer Learning}} & $s=64$& \mOc{85.1} & 92.5 & \mOc{86.1} & {\bf90.6} & \mOc{86.1} & {\bf90.6} & \mOc{86.1} & {\bf90.6} & {\bf167}--180\\
%\cline{2-12} 
\multicolumn{1}{|c|}{ImageNet\,$\to$\,CIFAR} & $s=128$ & \mOc{84.9} & {\bf86.7} & \mOc{85.6} & 90.6 & \mOc{85.5} & 90.6 & \mOc{85.5} & 90.6 & {\bf302}--332\\
\hline
\end{tabular}
%}
\end{table*}

\begin{table*}[]
\caption{Accuracy \TA, watermark recovery rate \rec\ with JPEG Compression, and the corresponding range of rarity in bits}
\label{tab:JPEGresults}
\centering
%\resizebox{\textwidth}{!}{%
%\begin{tabular}{|cl||cl|cl|cl|cl|cl|}
\begin{tabular}{|cl||cc|cc|cc|cc|cc||c|}
\hline
\multicolumn{2}{|c||}{} & \mTc{\textbf{QF=55}} & \mTc{\textbf{QF=65}} & \mTc{\textbf{QF=75}} & \mTc{\textbf{QF=85}} & \multicolumn{2}{c||}{\textbf{QF=95}} & \bf{Rarity}\\
%\cline{3-12} 
\multicolumn{2}{|c||}{\textbf{Dataset $\backslash$ Nb. triggers}} & \mOc{\TA} & \mOc{\rec} & \mOc{\TA} & \mOc{\rec} & \mOc{\TA} & \mOc{\rec} & \mOc{\TA} & \mOc{\rec} & \mOc{\TA} & \mOc{\rec} & \bf{$R$ in bits}\\
\hline

\multicolumn{1}{|c|}{} & $s=40$ & \mOc{99.3} & \bf{82.5} & \mOc{99.3} & \bf{82.5} & \mOc{99.3} & \bf{82.5} & \mOc{99.3} & \bf{82.5} & \mOc{99.3} & \bf{82.5} & {\bf{86}}--86\\
%\cline{2-12} 
\multicolumn{1}{|c|}{\textbf{MNIST}} & $s=64$ & \mOc{99.2} & \bf{89.1} & \mOc{99.2} & \bf{89.1} & \mOc{99.2} & \bf{89.1} & \mOc{99.2} & \bf{89.1} & \mOc{99.2} & \bf{89.1} & {\bf{167}}--167\\
%\cline{2-12} 
\multicolumn{1}{|c|}{} & $s=128$ & \mOc{99.1} & \bf{87.5} & \mOc{99.0} & 88.3 & \mOc{99.0} & \bf{87.5} & \mOc{99.1} & \bf{87.5} & \mOc{99.1} & 87.5 & {\bf{308}}-320\\
\hline

\multicolumn{1}{|c|}{} & $s=40$ & \mOc{91.0} & 85.0 & \mOc{91.3} & \bf{82.5} & \mOc{91.5} & 85.0 & \mOc{91.9}   & 85.0 & \mOc{91.8} & 85.0 & {\bf{86}}--91\\
%\cline{2-12} 
\multicolumn{1}{|c|}{\textbf{Fashion MNIST}} & $s=64$ & \mOc{91.3} & \bf{78.1} & \mOc{91.4} & 84.4 & \mOc{91.6} & 85.9 & \mOc{91.6} & 87.5 & \mOc{91.9} & 87.5 & {\bf{122}}--155\\
%\cline{2-12} 
\multicolumn{1}{|c|}{} & $s=128$ & \mOc{90.6} & \bf{83.6} & \mOc{91.1} & 87.5 & \mOc{91.3} & 88.3 & \mOc{91.5} & 90.6 & \mOc{91.8} & 90.6 & {\bf{285}}--332\\
\hline

\multicolumn{1}{|c|}{} & $s=40$ & \mOc{78.2} & \bf{82.5} & \mOc{78.9} & 85.0 & \mOc{80.4} & 85.0 & \mOc{81.4} & 90.0 & \mOc{82.6} & 92.5 & {\bf{86}}--110\\
%\cline{2-12} 
\multicolumn{1}{|c|}{\textbf{CIFAR10}}& $s=64$ & \mOc{78.5} & 65.6 & \mOc{79.1} & \bf{64.1} & \mOc{80.1} & 79.7 & \mOc{81.3} & 81.3 & \mOc{82.5} & 85.9 & {\bf{86}}--149\\
%\cline{2-12} 
\multicolumn{1}{|c|}{} & $s=128$ & \mOc{77.4} & \bf{57.8} & \mOc{78.9} & 64.8 & \mOc{79.7} & 75.8 & \mOc{81.2} & 78.9 & \mOc{82.7} & 85.1 & {\bf{131}}--290\\
\hline

\multicolumn{1}{|c|}{} & $s=40$ & \mOc{83.6} & \bf{87.5} & \mOc{84.9} & 92.5 & \mOc{84.7} & 90.0 & \mOc{85.3} & 92.5 & \mOc{85.7} & 92.5 & {\bf{97}}--110\\
%\cline{2-12} 
\multicolumn{1}{|c|}{\textbf{Transfer Learning}} & $s=64$& \mOc{83.9} & \bf{87.5} & \mOc{84.3} & \bf{87.5} & \mOc{85.0} & 89.1 & \mOc{85.6} & 90.6 & \mOc{85.9} & 90.6 & {\bf{155}}--167\\
%\cline{2-12} 
\multicolumn{1}{|c|}{ImageNet\,$\to$\,CIFAR} & $s=128$ & \mOc{83.1} & \bf{81.2} & \mOc{83.7} & 82.8 & \mOc{84.3} & 89.8 & \mOc{84.9} & 90.6 & \mOc{85.2} & 90.6 & {\bf{263}}--332 \\
\hline
\end{tabular}
%}
\end{table*}

\subsection{Robustness against Attacks}
\label{subsec:attacksperformance}

%We begin by looking at the fine-tuning attack and how it affects the proposed watermarking algorithm. We can see that the test accuracy on the original task before fine-tuning is preserved after the attack in all of the cases reported in the TA-FT column of Table \ref{tab:PerformanceSecureHash}. The transfer learning scenario with $N=128$, for example, has the highest drop in test accuracy after fine-tuning attack. When compared to the B-TA, the TA dropped by 0.0148 in this case.

The fine-tuning attack has no effect on the watermark recovery, as evidenced by the values reported in Table~\ref{tab:PerformanceSecureHash}. In the worst case, after the fine-tuning, \rec\ drops by 3.91 percentage points (transfer learning with $s=128$).
The results also show that the test accuracy and watermark recovery are unaffected by all attacks based on weight quantization.
In terms of test accuracy, CIFAR10 and transfer learning suffer the greatest loss: a drop of 0.64~pp for CIFAR10 with $N=64$, 0.9~pp for transfer learning with $N=128$ against Float16. The test accuracy of the MNIST and Fashion MNIST is almost unaffected by any of the quantization attacks.
In the same way, the watermark recovery rate shows no sign of degradation.

Table~\ref{tab:JPEGresults} presents the results of the JPEG attack. Starting with MNIST, we see that JPEG compression has almost no effect on test accuracy and watermark recovery. In the worst-case, with JPEG QF of 65 and $s=128$ triggers, the loss of \TA\ is 0.4~pp, while the watermark recovery rate is nearly unchanged. Fashion MNIST share the same properties as MNIST. For example, with a JPEG QF of $55$ and $s=64$ triggers, the biggest loss in \TA\ is 0.91~pp. The watermark recovery rate decreases the most by 9.4~pp with $s=64$ and 7.0~pp with $s=128$.
In both cases, a watermark recovery rate of at least 78\% guarantees a rarity $R$ well above 120 bits.

JPEG has a bigger impact on RGB datasets.
For CIFAR10, the loss of \rec\ can be as severe as 32.0 percentage points. Note that this is accompanied by a loss of 6.4 points in test accuracy compared to the original DNN model. Yet this can be compensated by a bigger number of triggers: for $s=128$, the rarity is still above 130 bits.
%In the case of deploying the DNN after his "fake" ownership claim, the reduction in test accuracy is a significant loss for the attacker. Even if he were to beat the true Owner's 57.81\% in front of a Judge, he would lose performance in test accuracy, which would make the model's performance unacceptable. 
Transfer learning is more robust in terms of \TA\ and \rec\ than starting from scratch on CIFAR10 because we start with a well-trained model on a similar dataset with a larger number of classes. This model has already learned more useful features for the task at hand.

Transfer learning provides an advantage as well when the Owner wants to distribute multiple copies of the model.
There is no need to train the model from scratch in this scenario because the embedding of the watermark happens during transfer learning.
In this scenario, a different set of triggers could be used for watermark injection for each user and his assigned model in order to distinguish each DNN user. This is especially true because we tested the injection with transfer learning for much fewer training epochs, i.e. 30 and 50, to speed up the model distribution process, and we found no noticeable loss in performance for all the attacks.
However, this is not possible in the other cases for shallower models according to our experiment.

\subsection{Pruning}

\begin{figure}[b]
\centerline{\includegraphics[width=0.8\columnwidth]{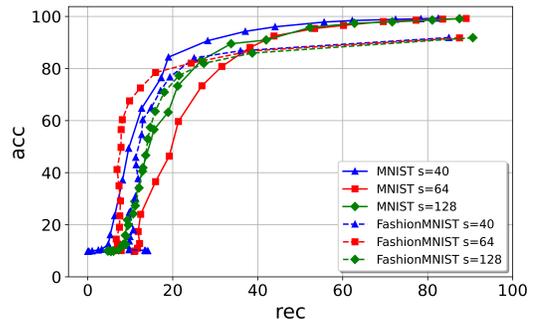}}
\caption{Test accuracy \TA\ vs. watermark recovery rate \rec\ for MNIST and Fashion MNIST with global pruning}
\label{fig:pruningMNISTANDFashionMNISTSecureHashTAvsWMRec}
\end{figure}

\begin{figure}[bt]
\centerline{\includegraphics[width=0.8\columnwidth]{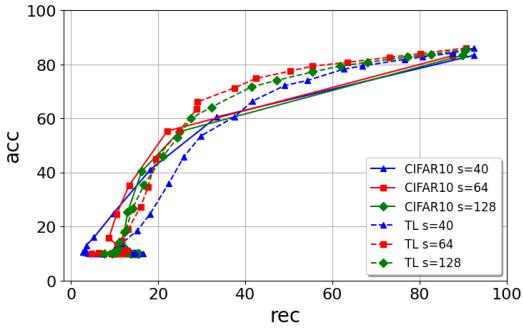}}
\caption{Test accuracy \TA\ vs. Watermark recovery rate \rec\ for CIFAR10 and Transfer Learning with global pruning}
\label{fig:pruningCIFAR10andImageNetSecureHashTAvsWMRec}
\end{figure}

\begin{figure}[bt]
\centerline{\includegraphics[width=0.8\columnwidth]{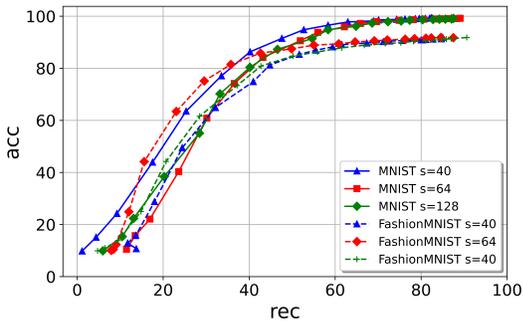}}
\caption{Test accuracy \TA\ vs. watermark recovery rate \rec\ for MNIST and Fashion MNIST with FC pruning}
\label{fig:FCpruningMNISTANDFashionMNISTSecureHashTAvsWMRec}
\end{figure}

\begin{figure}[bt]
\centerline{\includegraphics[width=0.8\columnwidth]{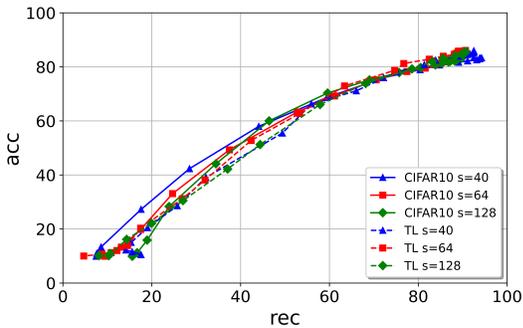}}
\caption{Test accuracy \TA\ vs. Watermark recovery rate \rec\ for CIFAR10 and Transfer Learning with FC pruning}
\label{fig:FCpruningCIFAR10andImageNetSecureHashTAvsWMRec}
\end{figure}

We now explore the limits of our scheme with an extreme pruning attack. In Fig.~\ref{fig:pruningMNISTANDFashionMNISTSecureHashTAvsWMRec},~\ref{fig:pruningCIFAR10andImageNetSecureHashTAvsWMRec},~\ref{fig:FCpruningMNISTANDFashionMNISTSecureHashTAvsWMRec}, and~\ref{fig:FCpruningCIFAR10andImageNetSecureHashTAvsWMRec}, the pruning is performed by randomly setting the weights of each layer with a pruning rate $k\in0.05*\{0,\ldots,19\}\cup\{ 0.97, 0.99\}$. For each point, the average performance of 50 pruning rounds is reported. Extreme pruning rates clearly remove the watermark. Yet, they also ruin the accuracy making the attacked model useless.    

Pruning globally the network harms more the recovery of the watermark than pruning only the fully connected layers. This is visible when comparing Fig.~\ref{fig:pruningMNISTANDFashionMNISTSecureHashTAvsWMRec} and~\ref{fig:pruningCIFAR10andImageNetSecureHashTAvsWMRec} to Fig.~\ref{fig:FCpruningMNISTANDFashionMNISTSecureHashTAvsWMRec} and~\ref{fig:FCpruningCIFAR10andImageNetSecureHashTAvsWMRec}.
This is explained by the fact that pruning is parametrized by a rate. At a given rate, pruning globally the network removes many more neurons than pruning only the fully connect layers.

%The surprise is that the test accuracy \TA\ drops suddenly when the pruning rate increases, while the watermark recovery rate \rec\ smoothly degrades. At $k=0.99$, the model achieves accuracy lower than 84\% on MNIST and 59\% on Fashion MNIST, which are very low numbers for these datasets. Yet, a slightly less severe pruning ($k=0.90$ for MNIST, 0.7 for Fashion MNIST) almost removes the watermark while not spoiling the accuracy. \teddy{Note that the number of triggers must be in the order of $35\times R$ if \rec\ equals 0.2 according to~\eqref{eq:sLinearR}.
%This means more than $700$ triggers for $R=20$ bits!}
%In the case of MNIST, we can see that as the pruning ratio $k$ increases, the watermark recovery degrades, but this success comes at the expense of a significant drop in test accuracy, as $k=0.99$, TA-Prun is approximately 83.7\% with SHA1, 71.51\% with SHA256, and 60.85\% with SHA512. In the right side of the same Figure, the same observations apply to the Fashion MNIST case. For example, TA-Prun scores 42.42\% with SHA1, 36.89\% with SHA256, and 58.91\% with SHA512 for $k=0.99$.

Deeper models have a different behavior than shallow models, see Fig.~\ref{fig:pruningCIFAR10andImageNetSecureHashTAvsWMRec} and~\ref{fig:FCpruningCIFAR10andImageNetSecureHashTAvsWMRec}.
%First, CIFAR10 and transfer learning perform no better than random guessing for the high pruning ratio $k$.
The test accuracy \TA\ and watermark recovery rate \rec\ degrades more smoothly with the pruning rate $k$.
%when an attacker uses fully connected pruning attacks on a transfer learning model, the TA and watermark recovery rates drop much faster than when the entire model weights are pruned.
This is due to the model's feature extraction part which has more weight redundancy than the newly stacked decision part.
Fully connected pruning attacks are much less advantageous to the attacker compared to pruning the entire model.
As a result, even when strong attacks are used, watermark recovery degradation does not occur without a loss in model accuracy on the primary task.
%, motivating the attacker to find a good trade-off between test accuracy loss and proofing his DNN ownership.

The crucial question is whether removing the watermark sufficiently degrades the model so that it becomes useless. Let us assume that the Verifier grants the ownership if the rarity $R$ is bigger than 20 bits. This means that a random key produce an accepted proof of ownership with a probability lower than one over one million. Or, in terms of work, the Usurper needs dozens of millions of hash computations to forge a convincing enough set of triggers. According to~\eqref{eq:BetaInc}, $R\geq 20$ implies that the watermark recovery rate must be bigger than 38\% for $s=40$, 32\% for $s=64$, and 25\% for $s=128$.
Pruning the fully connected layers at least cut the test accuracy by half for CIFAR10 and Transfer Learning, which is clearly unacceptable. For MNIST and FashionMNIST, this reduces the test accuracy down to $\lesssim 80\%$, which is also unacceptable for such easy classification problems. Global pruning is more harmful especially on MNIST and FashionMNIST (see Fig.~\ref{fig:pruningMNISTANDFashionMNISTSecureHashTAvsWMRec}).
In this case, increasing the number $s$ of triggers is the only solution: it has almost no impact on the characteristic $\TA=f(\rec)$ while it makes a big difference on the rarity. For $s=128$, $\TA$ is lower than $80\%$ when $\rec=25\%$.

Following the analysis of the various attacks, we can conclude that the proposed algorithm meets the robustness requirement, in the sense that any loss in watermark recovery is always accompanied with performance degradation on the original task test accuracy.
Furthermore, protecting the model with the proposed method requires the completion of the following tasks: hashing images to compute key labels at training time, and performing tests on a relatively few key trigger images at test time. All of the tasks are very light and have very short processing times.

\section{Conclusion and Future Works}\label{sec:conclusion}
This paper presents a new black-box DNN watermarking protocol that uses a cryptographic one-way hash function and the injection of key trigger-label pairs.
Its main features are that i) it does not impair the performance on the original task, ii) it allows to quantify the value of the proof of ownership and the security level, iii) it is robust against a wide range of attacks, based on the results of our experiments.
Future research directions are the security against more complex attacks, such as watermark overwriting, and
a  theoretical investigation of the relationship between the number of key trigger-label pairs, the learning capacity of the network, and the input space dimension to understand the watermarking algorithm's limitations and capabilities.

\section{Appendix}
\label{sec:App}
\subsubsection{MNIST and Fashion MNIST}
The network for MNIST is as follows: 1 conv. layer (64 filters); a max pooling; 1 conv. (128 filters); a max pooling; 2 f.c. layers (256 and 10 neurons). For all the conv. layers, the kernel size is 5 with ReLU activation.
The network is trained for 100 epochs with a batch size of 64. 
The CNN for the Fashion MNIST is the same as for MNIST except a dropout regularization of rate 0.2 after the last pooling layer and between the f.c. layers.

\subsubsection{CIFAR10 CNN}
The network for CIFAR10 is structured as follows: 2 conv. layers (32 filters); 2 conv. layers (64 filters); 2 conv. layers (128 filters); 2 conv. layers (256 filters) (each block of two conv. layers is followed by a $2 \times 2$ max-pooling and a dropout layer of rate 0.2); two f.c. layers (128 and 256 neurons) separated by a 0.2 dropout; final layer (10 neurons) with softmax activation. For all the conv. layers, the kernel size is $3$ with ReLU activation, initialized using He Uniform. The network is trained for 200 epochs with a batch size of 64.

\subsubsection{Transfer learning with ImageNet}
VGG19 pre-trained model on ImageNet is the base model for transfer learning. The network's decision part is replaced by two ReLU f.c. layers (1024 and 512 neurons) and the final layer (10 neurons) with softmax activation.
Transfer learning is performed over CIFAR10 for 200 epochs with a batch size of 64.
%The optimization algorithm is SGD with learning rate 0.001 and momentum 0.9.

During training, the cross entropy loss function is used in all cases. MNIST and Fashion MNIST use the Adam algorithm with default parameters, while CIFAR10 and transfer learning use SGD with learning rate 0.001 and momentum 0.9.

\bibliographystyle{ieeetr}
\bibliography{wifs.bib} 
\end{document}